\documentclass[aps,prl,twocolumn,superscriptaddress,showpacs]{revtex4-2}
\usepackage{graphicx}
\usepackage{mathrsfs}
\usepackage{bm}
\usepackage{amsmath}
\usepackage{amsfonts}
\usepackage{multirow}
\usepackage{bigstrut}
\usepackage{rotating}
\usepackage{booktabs}
\usepackage{color}
\usepackage{braket}
\usepackage{dcolumn}
\usepackage{enumerate}
\begin{document}

\title{Second-order topological corner states in zigzag graphene nanoflake with different types of edge magnetic configurations}

\author{Cheng-Ming Miao}
\affiliation{College of Physics, Hebei Normal University, Shijiazhuang 050024, China}

\author{Qing-Feng Sun}
\affiliation{International Center for Quantum Materials, School of Physics, Peking University, Beijing 100871, China}
\affiliation{Collaborative Innovation Center of Quantum Matter, Beijing 100871, China}
\affiliation{CAS Center for Excellence in Topological Quantum Computation, University of Chinese Academy of Sciences, Beijing 100190, China}

\author{Ying-Tao Zhang}
\email[]{zhangyt@mail.hebtu.edu.cn}
\affiliation{College of Physics, Hebei Normal University, Shijiazhuang 050024, China}

\begin{abstract}
  We study the energy spectrum and energy levels of the extended Kane-Mele model with magnetic atoms on their zigzag edges.
  It is demonstrated that the edges of ferromagnetism or antiferromagnetism are enough to break the time-reversal symmetry and host one-dimensional gapped edge states. Thus a second-order topological phase transition could happen, which leads to the emergence of topological in-gap zero-dimensional corner states. We also prove that the in-gap corner states are robust against corner defects and magnetic disorder. Our proposal based on the edge antiferromagnetism represents that the high-order topological states can be realized although the net magnetization of the material is zero. In addition, we discuss the influence of spin magnetization orientation on the degeneracy and energy of in-gap corner states.
\end{abstract}
\pacs{11.30.Er, 42.25.Bs, 72.10.-d}
\maketitle
\section{\uppercase\expandafter{\romannumeral 1}. Introduction}

Topological insulators (TIs) are new quantum states of matter that exhibit an ordinary insulator state in their interior but possess protected conducting states on their edge or surface \cite{Moore2010,Hasan2010,Qi2011,Bansil2016,Chiu2016,Ren2016}. The concept of TIs is beyond Landau$^{\prime}$s widely accepted classification of phases by spontaneous symmetry breaking and is typically characterized by topological invariants, which is a number that doesn$^{\prime}$t change under a smooth deformation of the Hamiltonian. There is an important concept in the field of TIs called ``bulk-boundary correspondence" \cite{Hasan2010,Teo2010,Essin2011,Isaev2011}, which refers to a one-to-one relationship between the number of topologically protected edge modes and the bulk topological invariant(\emph{e.g.} Chern number).  For example, there are ($z - 1$) dimensional gapless edge states for the $z$-dimensional TIs, and the number of gapless edge states corresponds exactly to the topological invariant of the $z$-dimensional TIs.

Recently, the concept of topology has been extended from first-order to higher-order \cite{Benalcazar2017}. Higher-order topological insulators (HOTIs) have new bulk-boundary correspondence \cite{Benalcazar2017a,Song2017,Langbehn2017,Peng2017,Ezawa2018,Ezawa2018a,Schindler2018}, i.e., $z$-dimensional $k$th-order ($k\geq 2$) topological system corresponds to the existence of ($z - k$) dimensional edge states ($z\geq k$).
Compared with conventional topological phases, the simplest three-dimensional second-order TIs host three-dimensional gapped bulk states and gapless one-dimensional hinge modes.
Both the three-dimensional third-order topological insulators and the two-dimensional second-order TIs host zero-dimensional corner states.
Experimentally, zero-dimensional corner states have been realized in various systems such as electrical circuits \cite{Imhof2018,Peterson2018,Hofmann2019,Ezawa2019a,Ezawa2019}, acoustic \cite{Ni2018,Xue2018,Xue2019,Weiner2020}, photonic crystals \cite{Serra-Garcia2018,Li2019,Xie2019,Chen2019,Kim2020,Iwamoto2021}, mechanical \cite{Rocklin2016,Wakao2020}, and ferromagnetic resonance \cite{Plekhanov2019} $et~al$.

Even though the research on HOTIs is progressing rapidly, the electron corner states in two-dimensional second-order TIs have not been observed experimentally. The important characteristic of a two-dimensional second-order TIs is the emergence of zero-dimensional in-gap corner modes \cite{addref1}. Based on this characteristic, several works pointed out that the alternating structure of strong and weak hoppings can induce the topological transition and then give rise to the appearance of zero-dimensional in-gap corner states \cite{Mizoguchi2019,Mizoguchi2020,Diop2020,Ezawa2021,Wang2021,Xue2021,Yu2021}.
Their results can be understood as an extension of the one-dimensional Su-Schrieffer-Heeger model \cite{Su1979}. The disadvantage of their proposal is that the corner states of the HOTIs depend heavily on the boundary configurations.
Subsequently, some symmetry including inverse symmetry $\hat{\mathcal{I}}$ \cite{Khalaf2018,Xiong2020}, mirror-reflection symmetry $\hat{\mathcal{M}}$ \cite{Langbehn2017,Schindler2018} and rotation symmetry $\hat{\mathcal{C}}$ \cite{Song2017,Schindler2018,Khalaf2018a,Qian2022} can be employed to systematically generate examples of HOTIs. The above studies show that the symmetry-protected corner states can exist stably even if the spatial symmetry is broken. In summary, one needs two steps for generating the HOTIs. First, it is necessary to destroy the symmetry protecting the first-order topological state, such as the time-reversal symmetry. Secondly, the way of breaking symmetry needs to have the various characteristics of anisotropy. Thus there are two options to accomplish these two steps. One is that a Wilson mass term is induced in Bernevig-Hughes-Zhang model \cite{Bernevig2006}, which will break the time-reversal symmetry and give rise to the opposite sign of the Dirac mass terms at the adjacent boundaries, then obtain the topologically protected corner states \cite{Yan2019,Ghosh2020}.
The other one is that by applying an in-plane magnetic field, the gapless edge states are gapped, which induces the edge topological transitions and then gives rise to the
zero-dimensional in-gap corner modes \cite{Ezawa2018b,Sheng2019,Ren2020,Han2022}.

In principle, the topological in-gap corner states can be generated when the gapless edge states are gapped. However, in the previous proposals,
an external magnetic field has been applied in the whole sample plane, which seems to be unnecessary.
In addition, it is demonstrated that the edge magnetism is
a robust property of zigzag graphene nanoribbons \cite{Son2006, Jung2009, Golor2013, Magda2014}.
Thus we predict that the topological in-gap corner states should emerge
by only considering the edge magnetism of graphene nanoribbons.
Compared to the ferromagnetism, antiferromagnetism is a magnetic
ordered state in which the magnetic moments are arranged in an antiparallel staggered order and do not exhibit macroscopic strong net magnetic moments.
The flip-flop magnetic moments in half of the sample will change the sample from uniform magnetization to staggered magnetization.
The antiferromagnetic order is widely available and can also break the time-reversal symmetry \cite{Tang2016,Lin2020}.
According to the above analysis, in this work, we induce the ferromagnetic and antiferromagnetic orders only at the boundary of Kane-Mele \cite{Kane2005} honeycomb lattice nanoflake. We investigate the energy spectrum of the modified Kane-Mele model with edge magnetic atoms.
The gapless edge states are gapped in both edge ferromagnetism and antiferromagnetism.
It is shown that the topological in-gap corner states appear in different positions of the diamond-shaped nanoflake, which have different energy degeneracy in various antiferromagnetic cases.
With the change of magnetization direction, the topological in-gap corner states always exist with different Fermi energy levels.
Finally, we also calculate the energy levels of the diamond-shaped nanoflake in which the magnetization directions at the four boundaries are different from each other. The results show that the energy values of the four corner states are no longer degeneracy, and the four separate in-gap corner states occupy four different corners of the diamond-shaped nanoflake respectively.

The rest of the paper is organized as follows. In Sec. II, we derive the tight-binding Hamiltonian for the modified Kane-Mele model with edge magnetic atoms.
In Sec. III, we show the band structure and energy levels of the system, and investigate the corner state distribution of a diamond-shaped open boundary system in different ferromagnetic and antiferromagnetic cases.
Finally, a summary is presented in Sec. IV.

\section{\uppercase\expandafter{\romannumeral 2}. model and hamiltonian}
\begin{figure}
	\centering
	\includegraphics[width=8cm,angle=0]{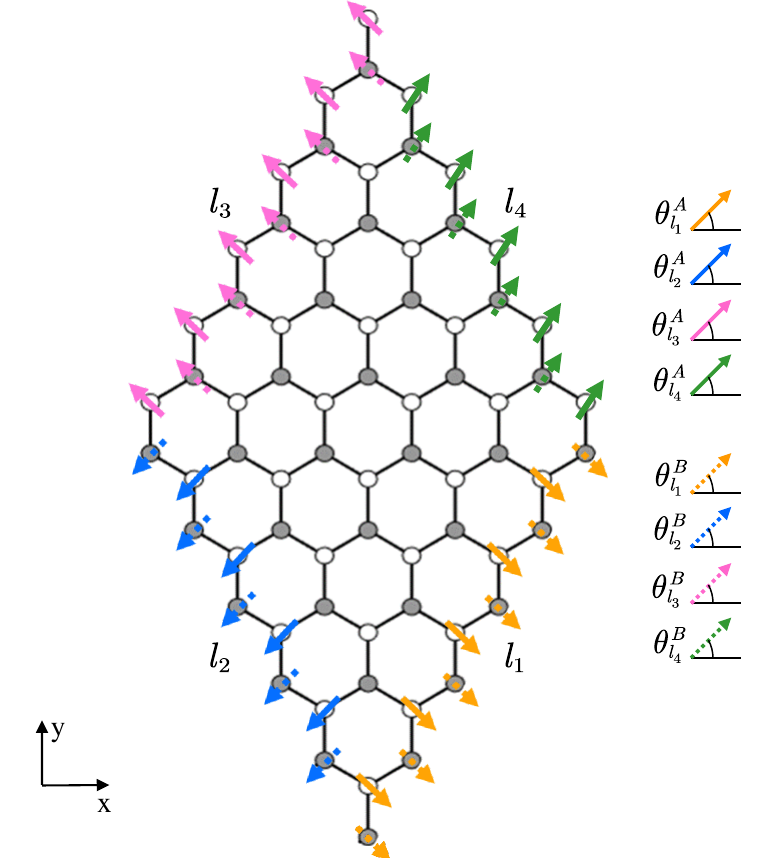}
	\caption{Schematic of diamond-shaped honeycomb lattice nanoflake with zigzag boundaries. Sublattices $A$ and $B$ are denoted by empty and full dots, respectively. The magnetization directions of magnetic atoms at zigzag boundaries $l_n$ ($n=1,2,3,4$) are indicated by yellow, blue, pink and green arrows respectively. The solid and dotted arrows indicate the spin magnetization orientation of the sublattices $A$ and $B$, respectively.}
	\label{fig1}
\end{figure}

We consider the modified Kane-Mele model in the presence of intrinsic spin-orbit coupling and edge magnetic atoms, as shown in Fig.~\ref{fig1}, where the four zigzag edges of a diamond-shaped honeycomb lattice nanoflake are labeled as $l_1$, $l_2$, $l_3$ and $l_4$. Each of the four zigzag edges has magnetic atoms with tunable magnetization directions, which are indicated by yellow, blue, pink, and green arrows respectively. The directions of magnetization at sublattices $A$ and $B$ are represented by solid line and dotted line arrows respectively. $\theta_{l_n}^{A}$ and $\theta_{l_n}^{B}$ are the angles between the magnetization vector $\mathbf{m}$ and the $x$ axis at sublattices $A$ and $B$ of zigzag boundary $l_n$.

The tight-binding Hamiltonian of this model can be written as follows \cite{Kane2005}:
 \begin{align}
  H=&-t\sum_{\left< i,j \right> ,\sigma}{c_{i\sigma}^{\dag}c_{j\sigma}}+\sum_{\left< \left< i,j \right> \right> ,\sigma ,\sigma'}{i \lambda \nu _{ij} c_{i\sigma}^{\dag}c_{j\sigma'}\left[ \mathbf{\hat{s}}_z \right] _{\sigma \sigma'}}\nonumber \\
  &+\sum_{l,\sigma ,\sigma'}{Jc_{l\sigma}^{\dag}c_{l\sigma'}\left[ \mathbf{\hat{m}}\cdot \mathbf{\hat{s}} \right] _{\sigma \sigma'}},
  \label{eq1}
 \end{align}
 where $c_{i\sigma}$ and $c_{i\sigma}^{\dag}$ are the annihilation and creation operators for an electron at the discrete site $i$ that belongs to the sublattice $A$ or $B$ and carries spin $\sigma$. The first term is the nearest-neighbor hopping term, and $t$ is the hopping energy. The second term is the intrinsic spin-orbit interaction with an amplitude of $\lambda$, which connects the second nearest neighbor. It depends on clockwise ($\nu _{ij}=-1$) or counterclockwise ($\nu _{ij}=1$) hopping paths from site $j$ to $i$. Since edge magnetism is shown to be a robust property of zigzag graphene nanoribbons in Refs.\cite{Son2006, Jung2009, Golor2013, Magda2014}, the last term in Eq.~(\ref{eq1}) is the in-plane exchange field of magnetic atoms in four boundaries $l_n$ $(n=1, 2, 3, 4)$, as shown in Fig.~\ref{fig1}. It is a spin exchange coupling that introduces spin magnetization and breaks time-reversal symmetry. The orientation of magnetization is along the unit vector $\mathbf{\hat{m}}$ = (cos $\theta_{l_n}^{A/B}$, sin $\theta_{l_n}^{A/B}$), where $\theta_{l_n}^{A/B}$ is the spin magnetization orientation angle on sublattice $A/B$ between $\mathbf{\hat{m}}$ and the $x$ axis at boundary $l_n$.
 $J=J_{A/B}$ stands for the exchange interaction energy at sublattice A/B.
 $\mathbf{\hat{s}}_{x,y,z}$ are Pauli matrices denoting spin space. Without loss of generality, we set $\lambda=0.1t,~J=J_{A/B}=0.2t$ in our calculations unless otherwise noted.

Here, we mainly study the various edge magnetic configurations caused by different magnetization directions.
We focus on the two cases where the magnetic directions of sublattices A and B are the same and opposite at the boundary.
Edge ferromagnetism means that sublattices A and B at the boundary
have the same magnetization direction ($\theta_{l_n}^{B}=\theta_{l_n}^{A}$).
The edge ferromagnetism in graphene can be induced by depositing ferromagnetic substrates at the boundaries of graphene nanoflake \cite{Haugen2008,Yang2013}, and the magnetization of the four boundaries can be individually controlled by the magnetic proximity effect of the four ferromagnetic substrates.
Edge antiferromagnetism is that the magnetization directions of sublattices A and B are opposite at the boundary ($\theta_{l_n}^{B}=\theta_{l_n}^{A}+\pi$), which is characteristic of the spontaneous magnetism of the graphene zigzag edge \cite{Son2006, Krompiewski2017}.
In addition, the magnetic moment is the largest at the zigzag terminals,
and the closer the lattice is to the center of the nanoflakes, the smaller its magnetic moment.
The exchange field could also be the original product of other physics, such as adsorbing atoms onto graphene or graphene-like nanoflakes to achieve atomic-scale control of graphene magnetism \cite{Gonzalez-Herrero2016,Weymann2020}.
It's worth noting that if the spin-orbit interaction term $\lambda$ is smaller than the magnitude of the magnetic moment of the bulk, the topologically nontrivial quantum spin Hall state is broken and there is no helical edge state \cite{Luo2020}. However, strong spin-orbit coupling is found in some graphene-like materials such as transition-metal dichalcogenides \cite{Taguchi2018}. Meanwhile, the magnetization of the lattices in the bulk region become negligible if the sample size of the nanoflake is significantly larger than the decay length of the spin-polarized edge states induced by the spontaneous magnetization \cite{Son2006}.

If all the magnetic atoms have the same spin magnetization direction, our proposed system exhibits ferromagnetism.
On the other hand, when total magnetization is zero, the system is antiferromagnetic. There are a variety of antiferromagnetic configurations.
In order to analyze the symmetry of different magnetic configurations,
we introduce the real-space mirror operations $\hat{\mathcal{M}}_{x/y}^{r}$ and spin-space mirror operations $\hat{\mathcal{M}}_{x/y}^{s}$ \cite{Yan2020,Cheng2021}.
The real-space mirror operations $\hat{\mathcal{M}}_{x}^{r}$ and $\hat{\mathcal{M}}_{y}^{r}$ act on the coordinates, that is, $\hat{\mathcal{M}}_{x}^{r}$ ($\hat{\mathcal{M}}_{y}^{r}$) turns $x$ ($y$) into $-x$ ($-y$),
and the spin-space mirror operations $\hat{\mathcal{M}}_{x}^{s}$ and $\hat{\mathcal{M}}_{y}^{s}$ act in the spin space, e.g., $\hat{\mathcal{M}}_{x}^{s}$ ($\hat{\mathcal{M}}_{y}^{s}$) changes $\hat{\mathbf{s}}_{x}$, $\hat{\mathbf{s}}_{y}$, $\hat{\mathbf{s}}_{z}$ to $\hat{\mathbf{s}}_{x}$, -$\hat{\mathbf{s}}_{y}$, -$\hat{\mathbf{s}}_{z}$
(-$\hat{\mathbf{s}}_{x}$, $\hat{\mathbf{s}}_{y}$, -$\hat{\mathbf{s}}_{z}$).
Based on the consideration of symmetry, we mainly consider three cases that are favorable for the appearance of in-gap corner states:
(i) The magnetization directions of four boundaries are the same ($\theta_{l_1}^{A/B}=\theta_{l_2}^{A/B}=\theta_{l_3}^{A/B}=\theta_{l_4}^{A/B}=0$), and the system is protected by mirror symmetry $\hat{\mathcal{M}}_{x}=\hat{\mathcal{M}}_{x}^{r} \otimes \hat{\mathcal{M}}_{x}^{s}$;
(ii) The magnetization directions of boundaries $l_1$ and $l_2$ are equal to those of $l_3$ and $l_4$ ($\theta_{l_1}^{A/B}=\theta_{l_3}^{A/B}=0,~\theta_{l_2}^{A/B}=\theta_{l_4}^{A/B}=\pi$), respectively. Mirror symmetry $\hat{\mathcal{M}}_{y}=\hat{\mathcal{M}}_{y}^{r} \otimes \hat{\mathcal{M}}_{y}^{s}$  protects the system;
(iii) The magnetization directions of boundaries $l_1$ and $l_2$ are the same as those of $l_4$ and $l_3$ ($\theta_{l_1}^{A/B}=\theta_{l_4}^{A/B}=0,~\theta_{l_2}^{A/B}=\theta_{l_3}^{A/B}=\pi$), respectively.
The system satisfies symmetry $\hat{\mathcal{M}}'_{x}=\hat{\mathcal{M}}_{x} \mathcal{T}$, in which $\mathcal{T}$ is time reversal symmetry.
 In addition to mirror symmetry, it also satisfies rotation symmetry $\hat{\mathcal{C}}_{2z}=\hat{\mathcal{C}}_{2z}^{r} \otimes \hat{\mathcal{C}}_{2z}^{s}$ and inversion symmetry $\hat{\mathcal{I}}'=\hat{\mathcal{I}} \mathcal{T}$, $\hat{\mathcal{I}}=\hat{\mathcal{I}}^{r} \otimes \hat{\mathcal{I}}^{s}$, in which $\hat{\mathcal{C}}_{2z}^{r}$ ($\hat{\mathcal{C}}_{2z}^{s}$) is the $\pi$ ratation around $z$ in real (spin)-space, and $\hat{\mathcal{I}}^{r}$ ($\hat{\mathcal{I}}^{s}$) is inversion in real (spin)-space. This is unique for case (iii), in contrast to cases (i) and (ii) where rotational symmetry $\hat{\mathcal{C}}'_{2z}=\hat{\mathcal{C}}_{2z} \mathcal{T}$ and inversion symmetry $\hat{\mathcal{I}}$ are satisfied. 
More abundant antiferromagnetic conditions can be obtained by the same or opposite magnetization orientation of magnetic atoms at the sublattices $A$ and $B$.

\section{\uppercase\expandafter{\romannumeral 3}.results and discussion}
\begin{figure}
	\centering
	\includegraphics[width=8.6cm,angle=0]{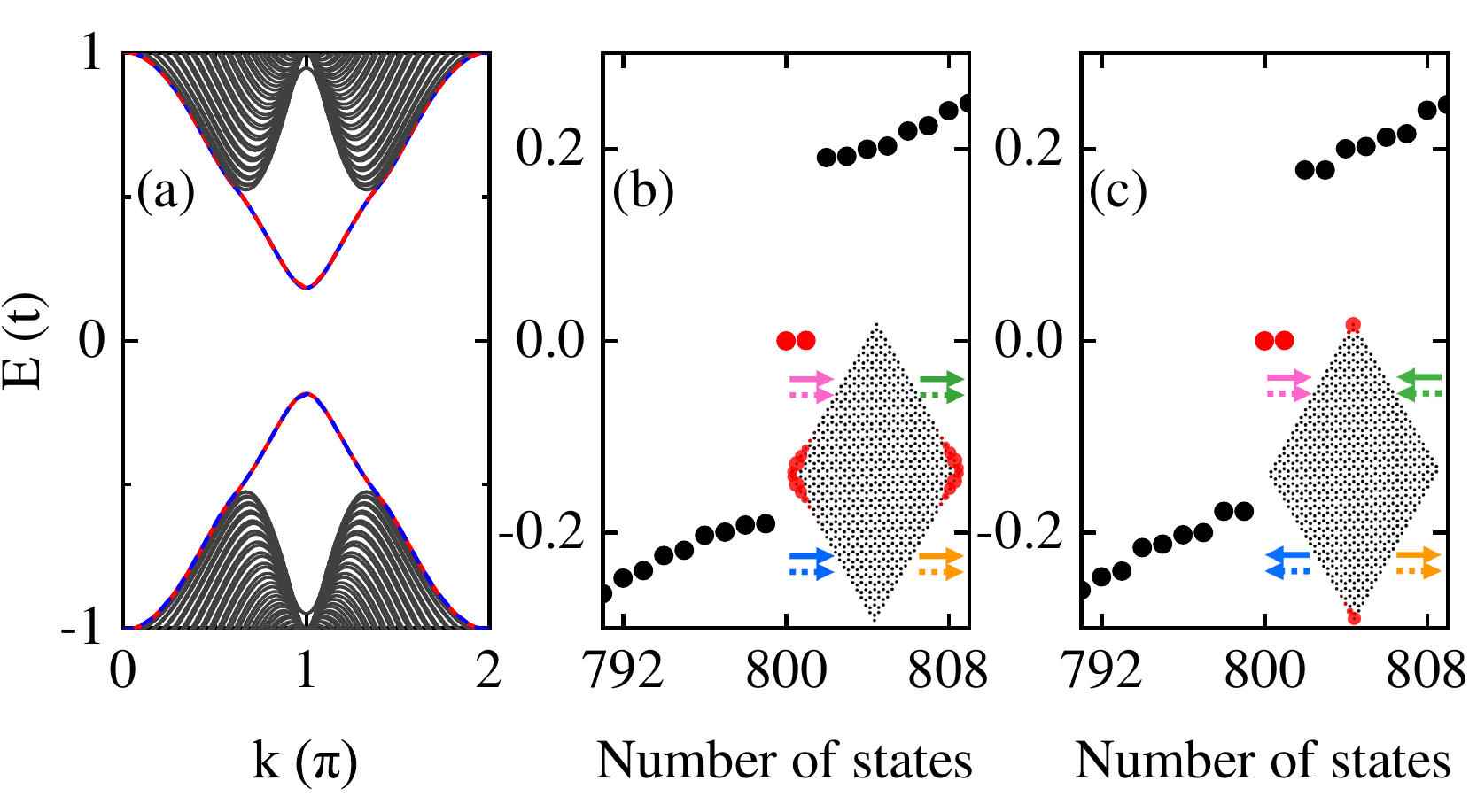}
	\caption{(a) Band structure for the zigzag nanoribbon with edge ferromagnetic atoms. The alternating red and blue lines represent the gapped edge states. Energy levels of diamond-shaped nanoflake with edge ferromagnetic atoms $\theta_{l_1}^{A}=\theta_{l_2}^{A}=\theta_{l_3}^{A}=\theta_{l_4}^{A}=0$ for (b), and antiferromagnetic atoms $\theta_{l_1}^{A}=\theta_{l_3}^{A}=0$, $\theta_{l_2}^{A}=\theta_{l_4}^{A}=\pi$ for (c). The black and red dots represent the gapped edge states and in-gap corner states, respectively. Probability distribution of the corner states is plotted in the inset. Colored solid line and dotted line arrows are magnetization directions of sublattices A and B at four zigzag boundaries, respectively. The other parameters are set to be $~\lambda=0.1t,~J=0.2t$, and $\theta_{l_n}^{B}=\theta_{l_n}^{A}$.}
	\label{fig2}
\end{figure}

In Fig.~\ref{fig2}(a), we calculate the band energy structure of the graphene nanoribbon with zigzag edge ferromagnetism. The parameters are set to be $t=1,~\lambda=0.1t,~J=0.2t,~\theta_{l_n}^{A}=\theta_{l_n}^{B}=0$.
It has been demonstrated that a pair of spin-helical gapless edge modes counterpropagate along a zigzag Kane-Mele nanoribbon \cite{Kane2005}.
In the presence of the edge ferromagnetic atoms, the spin-helical gapless edge states are gapped as illustrated by alternating red and blue lines in Fig.~\ref{fig2}(a).
The reason is that the time-reversal symmetry protecting the gapless edge states is broken by the introduction of ferromagnetic boundaries. Furthermore, we plot the energy levels of diamond-shaped nanoflakes with zigzag edge ferromagnetism in Fig.~\ref{fig2}(b).
In the calculation, we choose a typical diamond-shaped flake with length $L_n = 40a$, where $a$ is the lattice constant. The other parameters are the same as Fig.~\ref{fig2}(a). One can see that the zero-energy in-gap states arise as displayed by the red dots.
The probability distribution of a wave function at half-filling is highlighted in the inset of Fig.~\ref{fig2}(b), it can be seen that the corner modes with fractional charge $e/2$ are almost (due to finite system size) localized at the two obtuse angles of the diamond-shaped nanoflake. The physical mechanism behind is that in-plane ferromagnetism at graphene edges breaks the time-reversal symmetry and leads to boundary anisotropy, however, the mirror symmetry $\hat{\mathcal{M}}_x$ is preserved to protect in-gap corner states \cite{Ren2020}. Our results fully demonstrate that only considering the edge ferromagnetism is sufficient to induce the HOTIs and generate the topological in-gap corner states.

It is well known that antiferromagnets produce no stray fields and are insensitive to external magnetic field perturbations. Furthermore, the boundary with antiferromagnetic order has a variety of antiferromagnetic configurations and can also break the time-reversal symmetry.
In Fig.~\ref{fig2}(c), we plot the energy levels of edge antiferromagnetic system, where the magnetization directions of boundaries are set to be $\theta_{l_1}^{A}=\theta_{l_3}^{A}=0,~\theta_{l_2}^{A}=\theta_{l_4}^{A}=\pi,~\theta_{l_n}^{B}=\theta_{l_n}^{A}$.
Here we would like to point out that the energy structure of the graphene nanoribbon with zigzag edge antiferromagnetism is the same as that of graphene nanoribbon with zigzag edge ferromagnetism, and is not shown again. One can see that two zero-energy in-gap states emerge, which are represented by red dots in Fig.~\ref{fig2}(c). For two zero-energy in-gap states, the probability of their wavefunctions is highlighted in the inset of Fig.~\ref{fig2}(c). It can be seen that the $1/2$ electron charge is found to localize at acute corners and lead to the fractionalized charge distribution. Our results demonstrate that edge antiferromagnetism can generate topological in-gap corner states and realize the HOTIs also. Furthermore, one can see from Figs.~\ref{fig2}(b) and~\ref{fig2}(c) that corner states are protected by mirror symmetries $\hat{\mathcal{M}}_x$ and $\hat{\mathcal{M}}_y$, respectively. And corner states appear at the corners along the orientation of symmetry.

\begin{figure}
	\centering
	\includegraphics[width=8.6cm,angle=0]{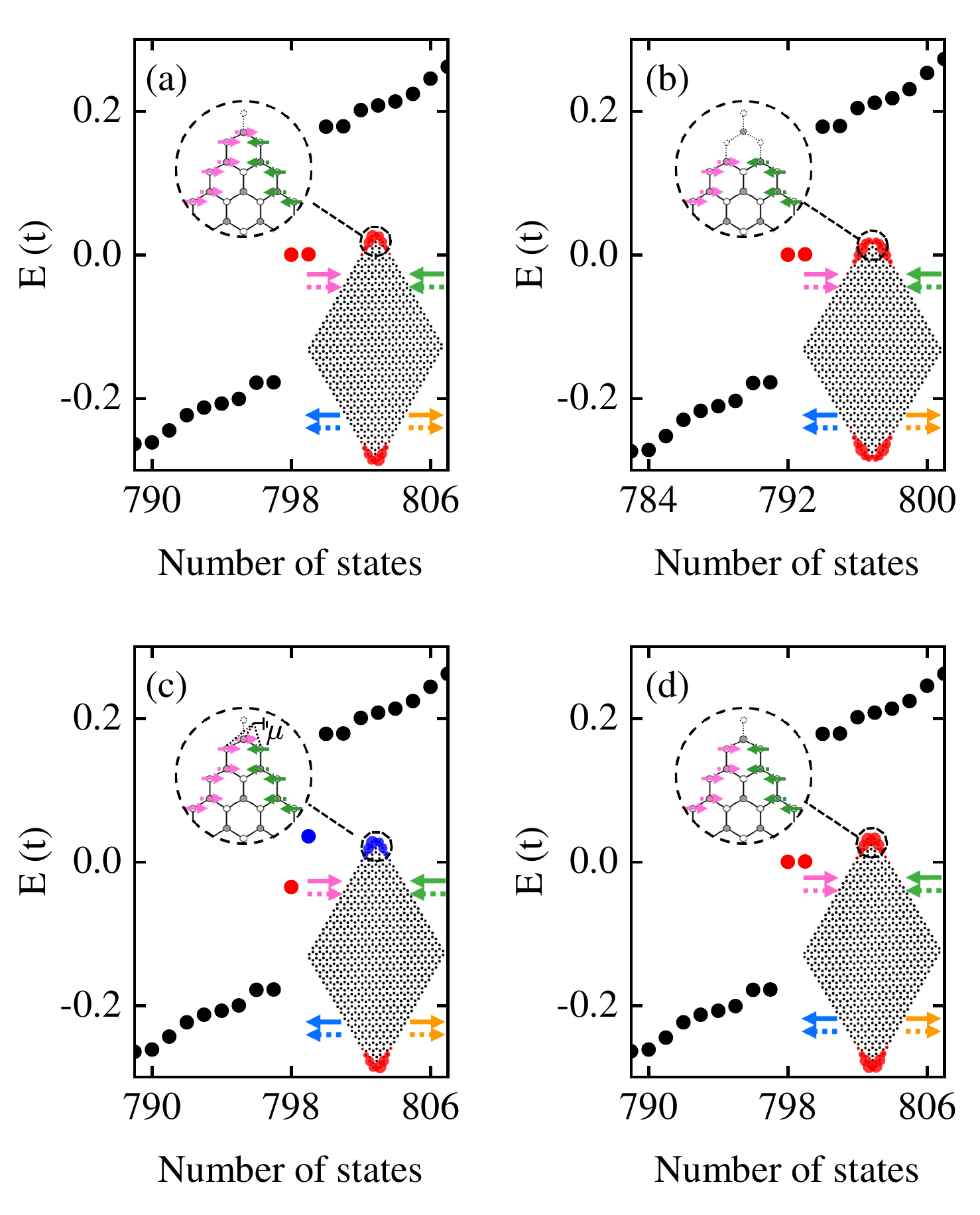}
	\caption{Energy levels of diamond-shaped nanoflake with edge antiferromagnetic atoms at different corner defects. Corner states are highlighted in red and blue. Probabilities of corner states are plotted in corresponding colors in the inset. Colored solid line and dotted line arrows are magnetization directions of sublattices A and B at four zigzag boundaries respectively. The details of the different corner defects are enlarged in the dotted circle, (a) one atom is vacant, (b) four atoms are vacant, (c) one atom is vacant and three atoms plus the on-site potential $\mu=0.1t$, (d) one atom is vacant and the one atom is not magnetic. The dotted lines show the atoms being removed. The upper and lower acute corner defects of the diamond-shaped nanoflake are the same except (c) three atoms at the upper (lower) acute angle plus on-site potential $\mu~(-\mu)$. Other parameters are the same as those in Fig.~\ref{fig2}(c).}
	\label{fig3}
\end{figure}

To verify that the corner states induced are robust, we plot the energy levels of diamond-shaped nanoflake with edge antiferromagnetic atoms at four different corner defects in Fig.~\ref{fig3}.
In Fig.~\ref{fig3}(a), we consider one atom is vacant at each of the two acute corners of the diamond-shaped nanoflake, as shown in the dotted circle in Fig.~\ref{fig3}(a). One can see that the edge states are gapped, and the in-gap zero-energy double degenerate corner states emerge, as represented by red dots in Fig.~\ref{fig3}(a). The wave function of the corner state is mainly distributed in the acute angle of the diamond-shaped nanoflake in the inset of Fig.~\ref{fig3}(a), indicating the corner states are robust against the atomic vacancy. In Fig.~\ref{fig3}(b), four atoms are removed at each of the two acute corners of the diamond-shaped nanoflake. It can be seen that the in-gap zero-energy double degenerate corner states still exist at the acute angle of the diamond-shaped nanoflake. Furthermore, the on-site potential terms $\mu$ are added to the three atoms of the upper acute angle and the lower acute angle of the diamond-shaped nanoflake, as shown in the dotted circle of Fig.~\ref{fig3}(c). Two corner states with different energy values are induced and represented by red and blue dots in Fig.~\ref{fig3}(c). The inset of Fig.~\ref{fig3}(c) shows that each of the two corner states is located on a spatial corner position of the diamond-shaped honeycomb lattice model. Lastly, the corner defect is set as one atom is removed and one atom is not magnetized at each of the two acute corners of the diamond-shaped nanoflake in Fig.~\ref{fig3}(d). The results show that the in-gap corner states won't disappear. The above results show that
the in-gap corners states still exist stably even if there are various defects in the corners, demonstrating that the in-gap corner states induced by edge (anti)ferromagnetism are topological and robust against corner defects.

\begin{figure}
	\centering
	\includegraphics[width=8.6cm,angle=0]{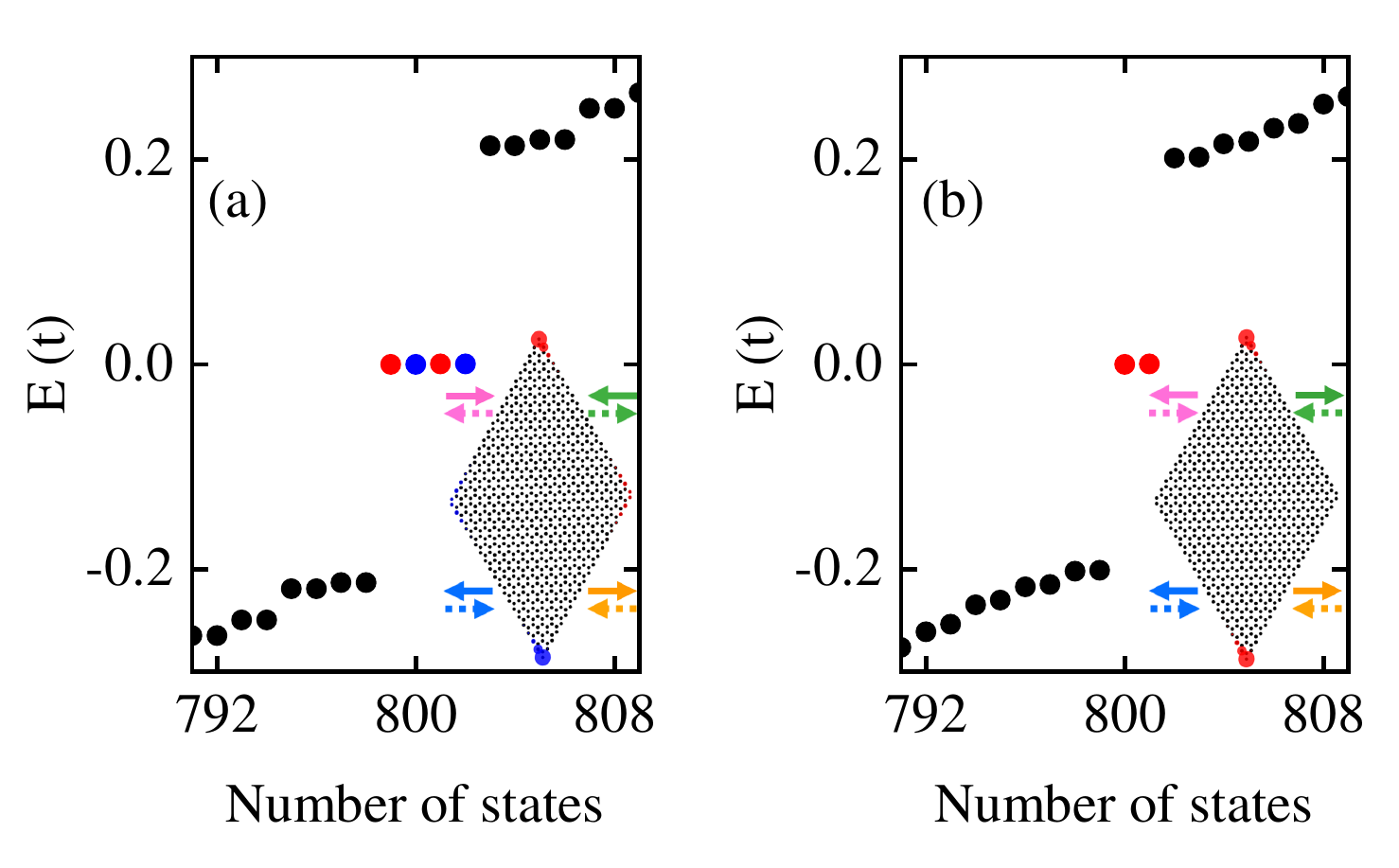}
	\caption{ Energy levels of diamond-shaped nanoflake with different edge antiferromagnetic atoms with opposite magnetization directions at sublattices $A$ and $B$ ($\theta_{l_n}^{B}=\theta_{l_n}^{A}+\pi$). (a) $\theta_{l_1}^{A}=\theta_{l_3}^{A}=0$, $\theta_{l_2}^{A}=\theta_{l_4}^{A}=\pi$, (b) $\theta_{l_1}^{A}=\theta_{l_4}^{A}=0$, $\theta_{l_2}^{A}=\theta_{l_3}^{A}=\pi$. Corner states are highlighted in red and blue. Probabilities of corner states are plotted in corresponding colors in the inset. Colored solid line and dotted line arrows are magnetization directions of sublattices A and B at four zigzag boundaries respectively. Other parameters are the same as those in Fig.~\ref{fig2}.}
	\label{fig4}
\end{figure}

Despite the entire diamond-shaped nanoflake sample behaves antiferromagnetic and has zero net magnetization in Fig.~\ref{fig2}(c), each of four boundaries are ferromagnetic configurations with $\theta_{l_n}^{B}=\theta_{l_n}^{A}$.
In Fig.~\ref{fig4}, we consider the neighboring magnetic moments point in opposite directions, i.e., $\theta_{l_n}^{B}=\theta_{l_n}^{A}+\pi$, which results in zero net magnetization of the each of edge and the entire sample.
In Fig.~\ref{fig4}(a), we set the parameters to be $\theta_{l_1}^{A}=\theta_{l_3}^{A}=0,~\theta_{l_2}^{A}=\theta_{l_4}^{A}=\pi,~\theta_{l_n}^{B}=\theta_{l_n}^{A}+\pi$ and calculate the energy levels of diamond-shaped honeycomb lattice nanoflake zigzag boundaries. It can be seen that the four-fold degeneracy zero-energy in-gap corner states emerge, which are indicated by red and blue dots in Fig.~\ref{fig4}(a). This indicates that the zero-energy in-gap corner states can be induced by antiferromagnetism even for the zero net magnetization edge. For every corner state, the probability of wavefunctions is highlighted in the inset of Fig.~\ref{fig4}(a) by its corresponding color. It is shown that two pairs of degenerate energy states are distributed on two different corners, respectively, which are protected by symmetry $\hat{\mathcal{M}}'_{x}$. By applying the different configuration parameters $\theta_{l_1}^{A}=\theta_{l_4}^{A}=0,~\theta_{l_2}^{A}=\theta_{l_3}^{A}=\pi,~\theta_{l_n}^{B}=\theta_{l_n}^{A}+\pi$, in Fig.~\ref{fig4}(b) we present the energy levels of diamond-shaped graphene nanoflake. One can see that there are two degenerate zero energy in-gap states in the discrete energy spectrum, as shown by two red dots.
The probability of their wavefunctions is highlighted in the inset of Fig.~\ref{fig4}(b). It can be seen that the $1/2$ electron charge spatial distribution is well located at the acute corners of the flake. These results demonstrate that even for the boundary of zero net magnetism, the zero-energy in-gap corner states can be induced by antiferromagnetism configuration, and the degeneracy of in-gap zero energy corner states can be adjusted by the spin magnetization orientation of antiferromagnetism.

\begin{figure}
	\centering
	\includegraphics[width=8.6cm,angle=0]{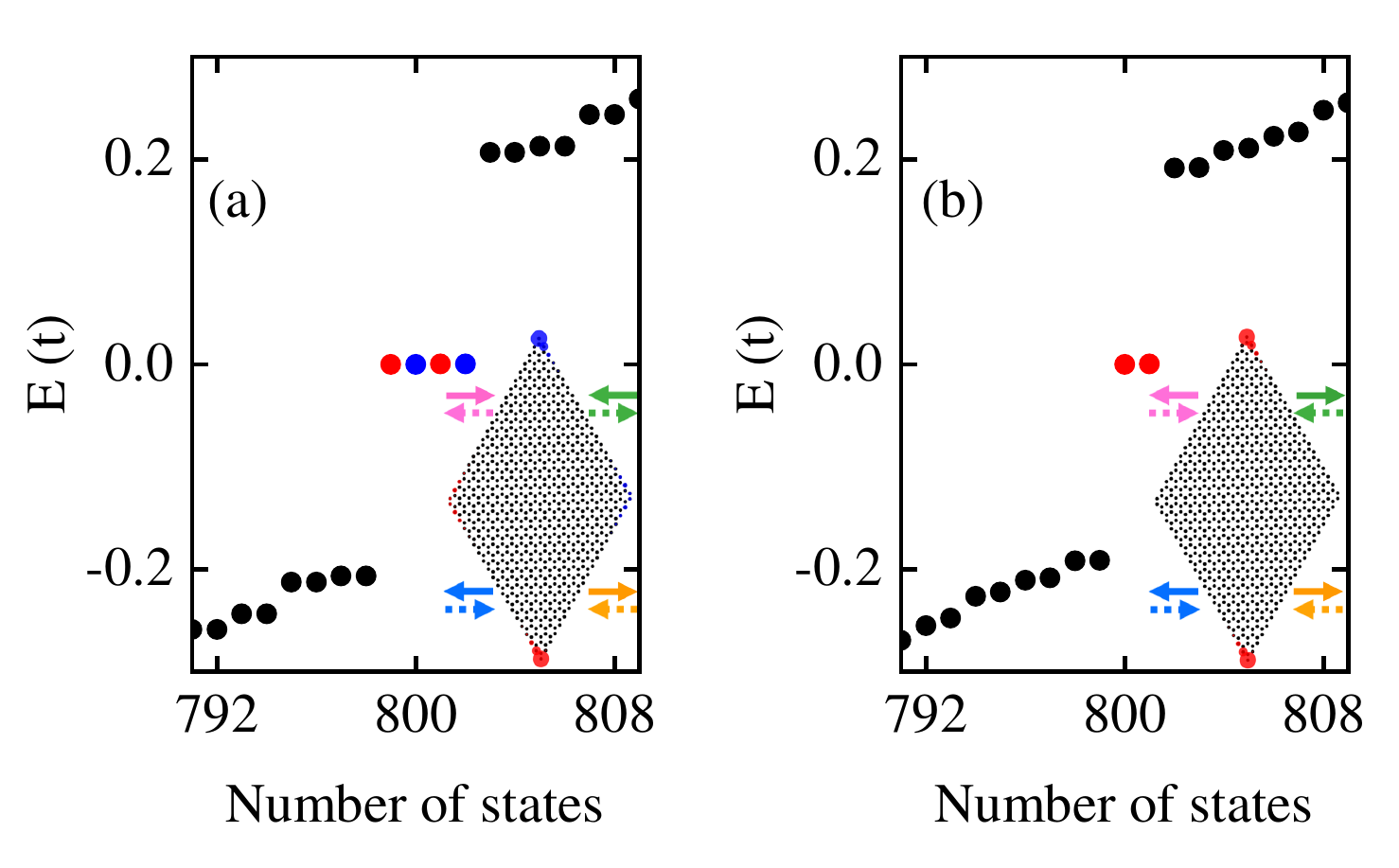}
	\caption{Energy levels of diamond-shaped nanoflake with different edge antiferromagnetic atoms with opposite magnetization directions and different exchange energies at sublattices $A$ and $B$ ($\theta_{l_n}^{B}=\theta_{l_n}^{A}+\pi$, $J_B=J=0.2t, J_A=0.1J_B$ for boundaries $l_1$ and $l_2$, $J_A=J=0.2t, J_B=0.1J_A$ for boundaries $l_3$ and $l_4$). (a) $\theta_{l_1}^{A}=\theta_{l_3}^{A}=0$, $\theta_{l_2}^{A}=\theta_{l_4}^{A}=\pi$, (b) $\theta_{l_1}^{A}=\theta_{l_4}^{A}=0$, $\theta_{l_2}^{A}=\theta_{l_3}^{A}=\pi$. Corner states are highlighted in red and blue. Probabilities of corner states are plotted in corresponding colors in the inset. Colored solid line and dotted line arrows are magnetization directions of sublattices A and B at four zigzag boundaries respectively. Other parameters are the same as those in Fig.~\ref{fig2}.}
	\label{fig5}
\end{figure}

Actually, the magnetic moments induced by spontaneous magnetization in graphene have different magnitudes and opposite directions for lattices A and B at the zigzag boundary \cite{Son2006, Jung2009}.
In Fig.~\ref{fig5}, we consider the neighboring magnetic moments point in opposite directions and different magnitudes, which is closer to the spontaneously induced magnetic moment in graphene.
It can be clearly seen from Fig.~\ref{fig1} that the outermost lattice of boundaries $l_1$ and $l_2$ is sublattice B, and the outermost lattice of boundaries $l_3$ and $l_4$ is sublattice A. So the parameters are selected as follows: $\theta_{l_n}^{B}=\theta_{l_n}^{A}+\pi$, $J_B=J=0.2t, J_A=0.1J_B$ for boundaries $l_1$ and $l_2$, $J_A=J=0.2t, J_B=0.1J_A$ for boundaries $l_3$ and $l_4$ in Fig.~\ref{fig5}.
For $\theta_{l_1}^{A}=\theta_{l_3}^{A}=0$ and $\theta_{l_2}^{A}=\theta_{l_4}^{A}=\pi$,
the four-fold degenerate zero-energy in-gap corner states emerge still,
which are highlighted by red and blue dots in Fig.~\ref{fig5}(a).
From the inset of Fig.~\ref{fig5}(a), it can be seen that two pairs of degenerate zero-energy states are distributed on two acute corners of the diamond-shaped nanoflake, respectively. Compared with Fig.~\ref{fig4}(a), each corner state in Fig.~\ref{fig5}(a) is at a different acute corner, but for the whole, the four-fold degenerate zero-energy in-gap corner states have the same density distribution. On the other hand, for the magnetic configuration parameters $\theta_{l_1}^{A}=\theta_{l_4}^{A}=0$ and $\theta_{l_2}^{A}=\theta_{l_3}^{A}=\pi$,
we find two degenerate zero energy in-gap states as displayed in Fig.~\ref{fig5}(b). The inset plots the probability of two in-gap states where the wave function is uniformly distributed at two corners. These results show that the zero energy corner states can be maintained even though the magnitudes of the magnetic moments at the A and B sublattices are different.

\begin{figure}
	\centering
	\includegraphics[width=8.6cm,clip]{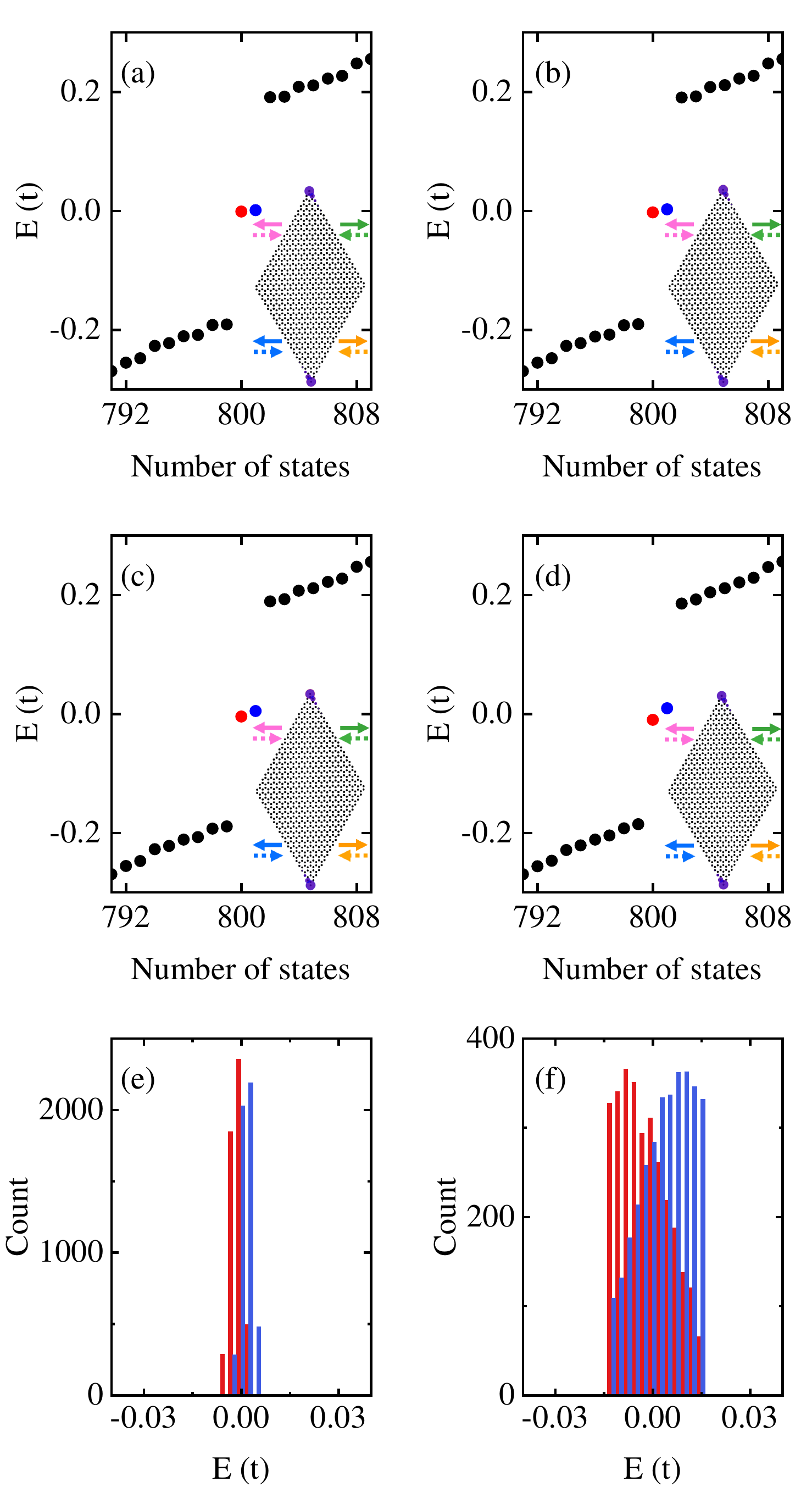}
	\caption{ (a)-(d) Energy levels of diamond-shaped nanoflake with different magnetic disorder strength $W=0.05$ for (a), $W=0.1$ for (b), $W=0.2$ for (c), and $W=0.4$ for (d).
	Corner states are highlighted in red and blue.
	Probabilities of corner states are plotted in the inset.
	Colored solid line and dotted line arrows are magnetization
	directions of sublattices A and B at four zigzag boundaries respectively.
 (e) (f) The energy distribution of two in-gap corner states with
 $W=0.05$ for (e) and $W=0.4$ for (f).
 Other parameters are the same as those in Fig.~\ref{fig5}(b).}
	\label{fig6}
\end{figure}

Here, we would like to point out that
the in-plane antiferromagnetic moments induced
by the spontaneous magnetism of the graphene zigzag edge
is the most advantageous configuration in the finite-size graphene-like system with intrinsic spin-orbit interaction, which has been demonstrated by several works applying the mean field theory calculations and quantum Monte Carlo simulations \cite{Weymann2016,Krompiewski2017,Raczkowski2020,Phung2020,Phung2022}. Their results have also shown that the magnitude of edge antiferromagnetic induced by spontaneous magnetization is inhomogeneous.
Thus below we calculate the energy levels and probability distribution
of diamond-shaped nanoflake with different magnetic disorder strength $W$ in Fig.~\ref{fig6}. In the presence of the magnetic disorder,
the disorder term $\sum_{l,\sigma ,\sigma'}{\omega J c_{l\sigma}^{\dag}c_{l\sigma'}\left[ \mathbf{\hat{m}'}\cdot \mathbf{\hat{s}} \right] _{\sigma \sigma'}}$ is added to Hamiltonian in Eq.~(\ref{eq1}), where $\omega$ is  randomly distributed in the interval [$0$, $W$] and
the unit vector $\mathbf{\hat{m}'}$ = (cos $\theta'$, sin $\theta'$)
with angle $\theta'$ being randomly distributed in the interval [$0$, $2\pi$].
All the curves are averaged over 5000 random configurations, which is enough to obtain reasonable results.
From Fig.~\ref{fig6}(a), one can see that the zero energy in-gap corner states still emerge at the magnetic disorder strength $W=0.05$, which are highlighted by red and blue dots. Each of the two corner states is uniformly distributed at two acute corners of the diamond-shaped nanoflake.
With the increasing of the disorder strength $W$, we can see that the energy values of the corner states are not exactly zero in Figs.~\ref{fig6}(b) and ~\ref{fig6}(c). When $W=0.4$, the degeneracy of in-gap states is completely broken. Even so, the in-gap states are still uniformly distributed at the two acute corners.
Figures ~\ref{fig6}(e) and ~\ref{fig6}(f) show that the energy value distribution of the two in-gap corner states with different magnetic disorder strength $W$.
The red and blue histograms of energy distribution in Figs.~\ref{fig6}(e) and ~\ref{fig6}(f) correspond to the in-gap states highlighted in red and blue [see Figs.~\ref{fig6}(a) and ~\ref{fig6}(d)], respectively.
In Fig.~\ref{fig6}(e), one can see that the energy values are clustered around zero with $W=0.05$.
When $W=0.4$, the energy values of the two corner states
are slightly diffused [see Fig.~\ref{fig6}(f)], but they still distribute
at a small range ($-0.015t$, $0.015t$) which is much smaller than
the edge gap ($-0.18t$, $0.18t$).
The above results show that the moderate magnetic fluctuations will not inhibit the emergence of in-gap corner states,
but affect the energy values and degeneracy of corner states.
It can be concluded that the in-gap corner states in our proposal
are robust against the effects of magnetic disorder.
Of course, if the magnetic moments are completely disordered at the boundary, the antiferromagnetic configuration is no longer stable, and the in-gap corner states will disappear.

\begin{figure}
	\centering
	\includegraphics[width=8.6cm,clip]{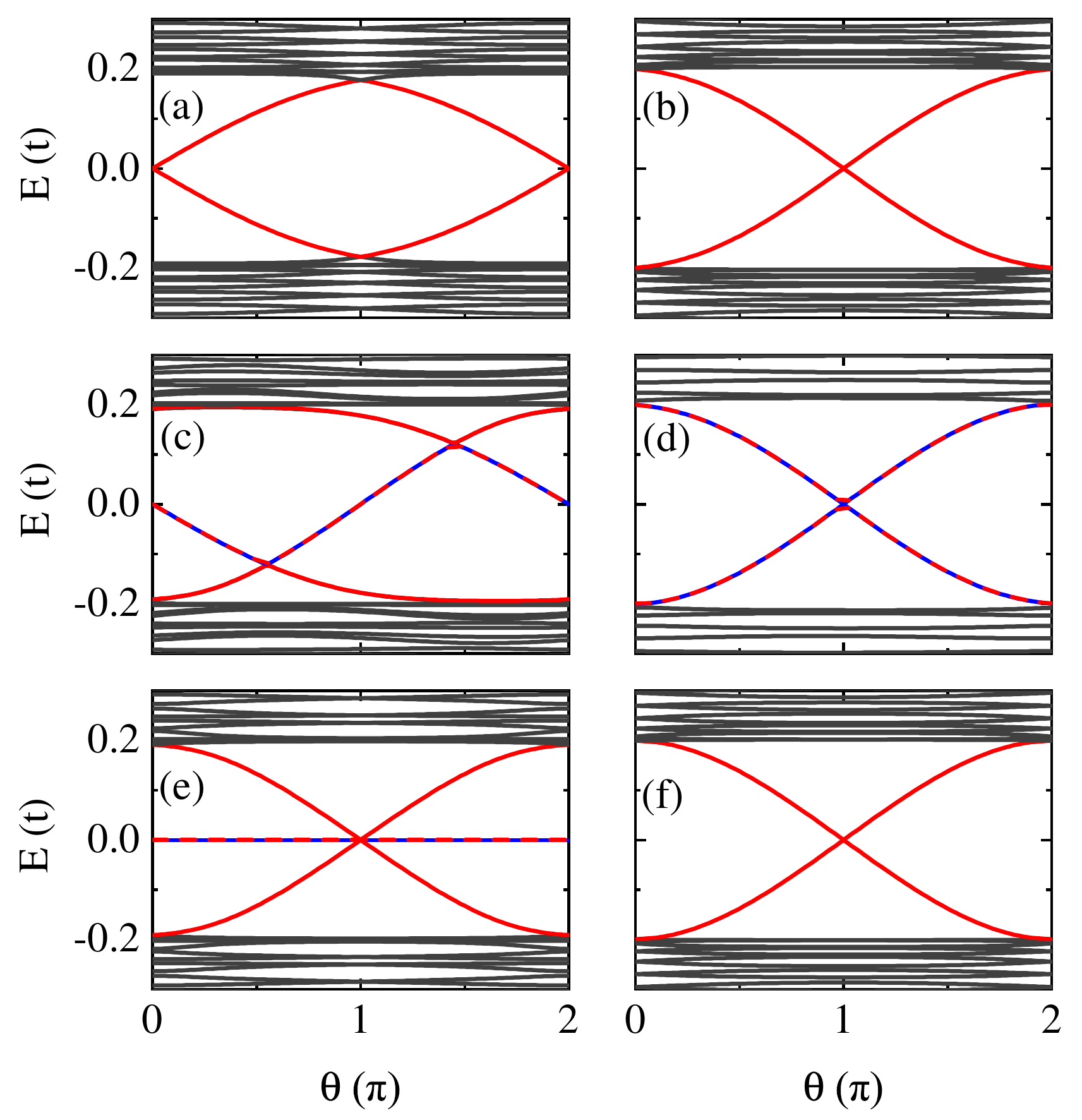}
	\caption{The energy spectrum of the diamond-shaped nanoflake versus angle $\theta$ between the magnetization vector $\mathbf{m}$ and the $x$ axis. Black lines denote the gapped edge states, red lines indicate non-degenerate corner states, and alternating red and blue lines represent double degenerate corner states. The parameters are taken to be $\theta_{l_1}^{A}=\theta_{l_2}^{A}=0$, $\theta_{l_3}^{A}=\theta_{l_4}^{A}=\theta$ in [(a), (b)], $\theta_{l_1}^{A}=\theta_{l_3}^{A}=0$, $\theta_{l_2}^{A}=\theta_{l_4}^{A}=\theta$ in [(c), (d)], and $\theta_{l_1}^{A}=\theta_{l_4}^{A}=0$, $\theta_{l_2}^{A}=\theta_{l_3}^{A}=\theta$ in [(e), (f)]. The magnetization directions of sublattices A and B are the same $\theta_{l_n}^{B}=\theta_{l_n}^{A}$ for (a), (c), (e) and opposite $\theta_{l_n}^{B}=\theta_{l_n}^{A}+\pi$ for (b), (d), (f). Other parameters are the same as those in Fig.~\ref{fig2}.}
	\label{fig7}
\end{figure}

To understand the effect of spin magnetization orientation
on the in-gap corner states.
In Fig.~\ref{fig7}, we plot the energy spectrum $E$ of the diamond-shaped graphene nanoflake as a function of spin magnetization orientation angle $\theta$ ($0\le \theta \le 2\pi$) with different (anti)ferromagnetic configurations. The concrete results and analysis of the different cases are below.

(i) In Fig.~\ref{fig7}(a), the parameters are set to be $\theta_{l_1}^{A/B}=\theta_{l_2}^{A/B}=0,~\theta_{l_3}^{A/B}=\theta_{l_4}^{A/B}=\theta$. The solid black lines in Fig.~\ref{fig7}(a) indicate the edge states, the gap of which is always open with the change of the angle $\theta$ from $0$ to $2\pi$. The red solid lines correspond to the in-gap corner states. One can see that the zero-energy two-fold degeneracy corner states exist with $\theta=0$, which become the corner states with non-zero-energy and non-degeneracy with the change of $\theta$ and completely disappear at $\theta = \pi$.

(ii) In Fig.~\ref{fig7}(b), we set the parameters to be $\theta_{l_1}^{A}=\theta_{l_2}^{A}=0,~\theta_{l_3}^{A}=\theta_{l_4}^{A}=\theta$. The spin magnetization orientations $\theta_{l_n}^{B}$ at sublattice $B$ are set to be opposite to $\theta_{l_n}^{A}$ at sublattice $A$, i.e., $\theta_{l_n}^{B}=\theta_{l_n}^{A}+\pi$. It is same as the result in Fig.~\ref{fig7}(a) that the gap of edge states is always open with the change of $\theta$. With the change of $\theta$ from $0$ to $2\pi$, two non-zero-energy and non-degeneracy corner states emerge, the energy values of which are symmetric with respect to $E=0$. At $\theta = \pi$ the zero-energy two degenerate corner states are formed.

(iii) In Fig.~\ref{fig7}(c), we fix the parameters as $\theta_{l_1}^{A}=\theta_{l_3}^{A}=0, ~\theta_{l_n}^{B}=\theta_{l_n}^{A}$ and set$ ~\theta_{l_2}^{A}=\theta_{l_4}^{A}=\theta$.  The alternating red and blue lines correspond to the in-gap two-fold degeneracy corner states. One can see that it is not symmetric about $\theta = \pi$. With the change of $\theta$ from $0$ to $2\pi$, triple degeneracy corner states appear with non-zero energy values. Double degenerate in-gap corner states with zero energy value forms  at $\theta =0$ and $\pi$.

(iv) In Fig.~\ref{fig7}(d), the parameters are set to be $\theta_{l_1}^{A}=\theta_{l_3}^{A}=0,~\theta_{l_n}^{B}=\theta_{l_n}^{A}+\pi$, and $~\theta_{l_2}^{A}=\theta_{l_4}^{A}=\theta$. It is shown that there are always two pairs of non-zero-energy double degenerate corner states, and the zero-energy quadruple degeneracy corner states form at $\theta =\pi$.

(v) In Fig.~\ref{fig7}(e), we choose to fix the parameters as $\theta_{l_1}^{A}=\theta_{l_4}^{A}=0,~\theta_{l_n}^{B}=\theta_{l_n}^{A}$ and set $\theta_{l_2}^{A}=\theta_{l_3}^{A}=\theta$. It is interesting that the zero-energy double degeneracy in-gap corner state is always present, and non-degeneracy in-gap corner states with non-zero-energy change in symmetry around $E=0$ with the change of $\theta$ from $0$ to $\pi$. Eventually a zero-energy quadruple degeneracy in-gap corner state forms at $\theta = \pi$.

(vi) In Fig.~\ref{fig7}(f), we present the result while the parameters are set to be $\theta_{l_1}^{A}=\theta_{l_4}^{A}=0,~\theta_{l_n}^{B}=\theta_{l_n}^{A}+\pi$ and $\theta_{l_2}^{A}=\theta_{l_3}^{A}=\theta$, the results in which are exactly the same as that in Fig.~\ref{fig7}(b).

The above calculation results and analysis show that the in-gap corner states can be adjusted by spin magnetization orientation in the diamond-shaped graphene system with various (anti)ferromagnetic edge configurations. The quadruple degenerate zero-energy corner states can form at $\theta=\pi$ in Fig.~\ref{fig7}(d) and ~\ref{fig7}(e). Triple degenerate corner states can appear in Fig.~\ref{fig7}(c). Thus quadruple degenerate, triple degenerate, double degenerate, and non-degenerate corner states can be controlled by the configurations of various (anti)ferromagnetic edges in our proposal.

\begin{figure}
	\centering
	\includegraphics[width=8.6cm,angle=0]{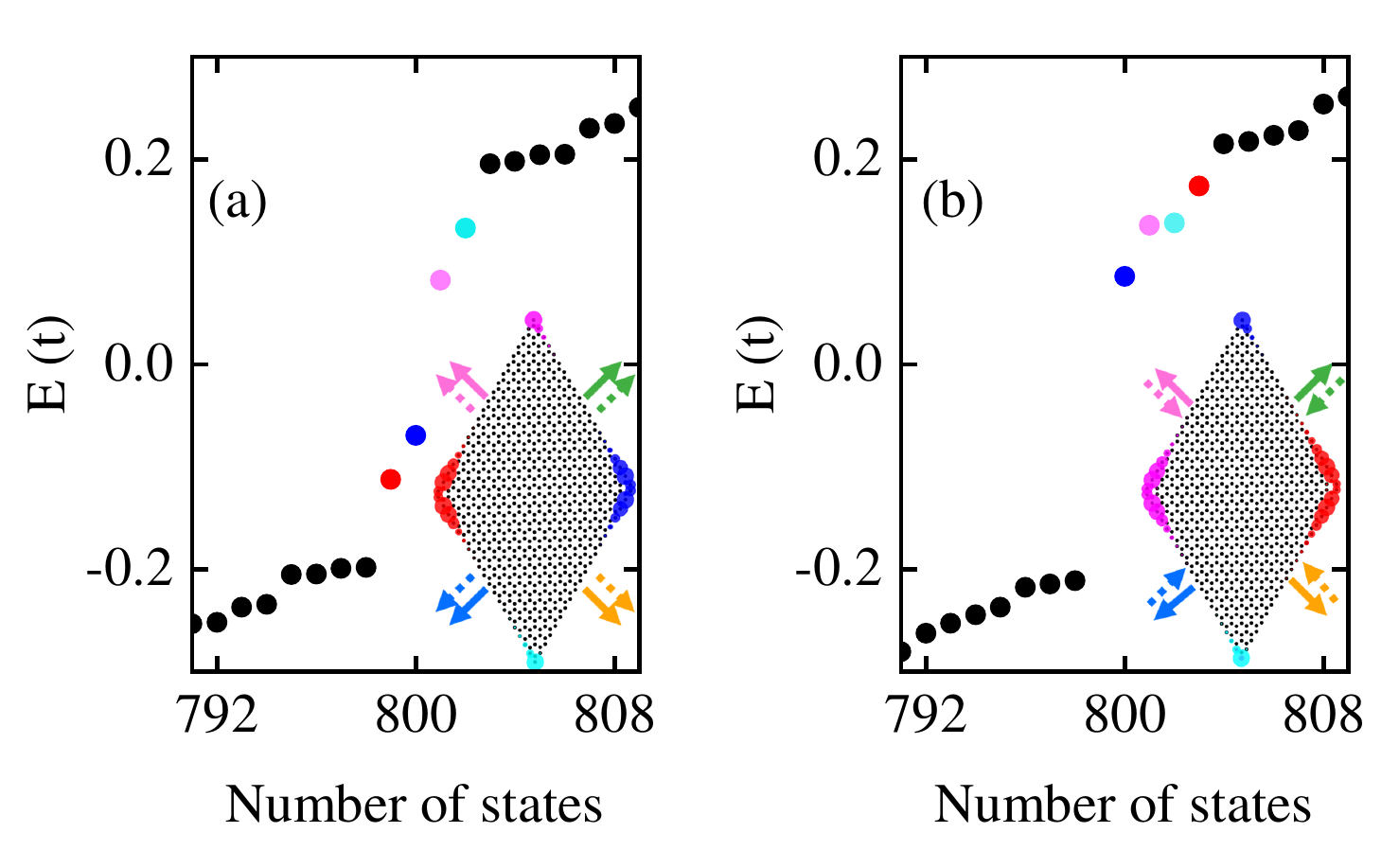}
	\caption{ Energy levels of diamond-shaped graphene nanoflake with edge antiferromagnetic atoms, which magnetization directions are different. Corner states with different eigenvalues are highlighted in red, blue, magenta and cyan. Probabilities of four corner states are plotted in corresponding colors in the inset. Colored solid line and dotted line arrows are magnetization directions of sublattices A and B at four zigzag boundaries respectively. The magnetization directions of sublattices A and B are the same $\theta_{l_n}^{B}=\theta_{l_n}^{A}$ for (a), and opposite $\theta_{l_n}^{B}=\theta_{l_n}^{A}+\pi$ for (b). The parameters are set to be $\theta_{l_1}^{A}=\pi/4,~\theta_{l_2}^{A}=3\pi/4,~\theta_{l_3}^{A}=5\pi/4,~\theta_{l_4}^{A}=7\pi/4$.}
	\label{fig8}
\end{figure}

Finally, we show the energy levels of diamond-shaped honeycomb lattice nanoflake with the four different boundaries magnetization directions in Fig.~\ref{fig8}, where the parameters are set to be $\theta_{l_1}^{A}=\pi/4,~\theta_{l_2}^{A}=3\pi/4,~\theta_{l_3}^{A}=5\pi/4,~\theta_{l_4}^{A}=7\pi/4$.
In both edge uniform magnetization ($\theta_{l_n}^{B}=\theta_{l_n}^{A}$) and staggered magnetization ($\theta_{l_n}^{B}=\theta_{l_n}^{A}+\pi$) cases, there are four corner states with non-degeneracy and non-zero-energy values, as shown by red, blue, magenta and cyan dots in Figs.~\ref{fig8}(a) and ~\ref{fig8}(b).
One can see from the inset of Fig.~\ref{fig8} that each of the corner states is located on a spatial corner position of the diamond-shaped honeycomb lattice model.
In Fig.~\ref{fig8}(a), it is shown that the energy levels of four in-gap corner states are up and down symmetrical around $E=0$,
the energy values of four in-gap corner states in Fig.~\ref{fig8}(b) are all above zero energy $E=0$.
\section{\uppercase\expandafter{\romannumeral 4}. summary}
In summary, we study the energy spectrum and energy levels of
the modified Kane-Mele model with zigzag edge ferromagnetism and antiferromagnetism.
Our results show that the zero-energy in-gap states arise
for the zigzag edge ferromagnetism,
which demonstrates that edge ferromagnetism is sufficient
to induce the higher-order topological insulators
and acquire in-gap corner states.
For the case of the zigzag edge antiferromagnetism,
we present that edge antiferromagnetism can also open the gap of gapless edge states, and induce the second-order topological insulators
and the zero-energy double degenerate corner states.
These results indicate that the second-order topological corner states
can be induced even in the zero net magnetism.
In particular, these in-gap corner states are robust against the corner defects and magnetic disorder.
Protected by different symmetries,
these corner states can show double degeneracy, triple degenerate,
quadruple degeneracy, and non-degeneracy states.
Furthermore, the spin magnetization orientation can control
the appearance and disappearance of corner states.
Because of a rich variety of the (anti)ferromagnetic order setting,
our proposal can well control the in-gap corner states
with different energy values and degeneracy.

\begin{acknowledgments}
\section{acknowledgments}
This work was financially supported by the National Natural Science Foundation of China (Grant No. 11474084, No. 12074097, and No. 11921005),
Natural Science Foundation of Hebei Province (Grant No. A2020205013),
and National Key R and D Program of China (Grant No. 2017YFA0303301).
\end{acknowledgments}


\end{document}